\documentclass[aps,prb,showpacs,twocolumn,groupedaddress,floatfix]{revtex4}

\usepackage{graphicx}% Include figure files
\usepackage{dcolumn}% Align table columns on decimal point
\usepackage{bm}% bold math

\begin{document}

%\preprint{}

\title{
Cold neutron scattering study on the diffuse and phonon
excitations in the relaxor Pb(Mg$_{1/3}$Nb$_{2/3}$)O$_{3}$}


\author{H. Hiraka$^{1,2}$, S.~-H.~Lee$^{3}$, P. M. Gehring$^{3}$, 
Guangyong Xu$^{1}$, and G. Shirane$^{1}$}
%\affiliation{
%Department of Physics, Brookhaven National Laboratory, Upton, New York 11973}
%\altaffiliation{ Institute for Materials Research, Tohoku University, 
%Sendai 980-8577, Japan}
%\author{S.~-H.~Lee and P. M. Gehring}
%\affiliation{NIST Center for Neutron Research, National Institute of Standards and Technology,
%Gaithersburg, Maryland 20899}

%\email[]{hiraka@bnl.gov}
%\thanks{}
%\altaffiliation{ Institute for Materials Research, Tohoku University, 
%Sendai 980-8577, Japan}
\affiliation{
   $^{1}$Department of Physics, Brookhaven National Laboratory, Upton, New York 11973 \\
   $^{2}$Institute for Materials Research, Tohoku University, Sendai 980-8577, Japan \\
   $^{3}$NIST Center for Neutron Research, National Institute of Standards and Technology,
   Gaithersburg, Maryland 20899}
%option in \documentclass). 
%\noaffiliation is required (may also be

\date{\today}

\begin{abstract}
%\begin{center}--- 18th edition ---
%\end{center}
Cold neutron scattering experiments have been performed to explore
the energy, temperature, and wave-vector dependence of the diffuse
scattering and the transverse acoustic (TA) phonons in the relaxor
Pb(Mg$_{1/3}$Nb$_{2/3}$)O$_{3}$.  We have observed a
weak, but definitive, diffuse scattering cross section above the
Burns temperature $T_{d}\sim 600$~K.  This cross section, which is
most likely caused by chemical short-range order, persists down to
100~K, and coexists with the much stronger diffuse scattering that
is attributed to the polar nanoregions.  A systematic study of the
TA phonon around $(1,1,0)$ has also been carried out.  The phonon
is well defined for small wave vectors $\mathbf{q}$, but broadens
markedly around $\mathbf{q}=(0.1, -0.1, 0)$.

\end{abstract}

%\pacs{PACS numbers: 75.30.Fv, 75.50.Ee, 75.40.Gb, 75.30.Ds}
%\keywords{}

\maketitle

%\section{}
%\subsection{}
%\subsubsection{}

\newpage

\section{INTRODUCTION}

The lead-oxide perovskite Pb(Mg$_{1/3}$Nb$_{2/3}$)O$_{3}$, or PMN,
is one of the most interesting and well-studied relaxor compounds
to date, primarily because of its exceptional piezoelectric
properties and enormous potential in industrial
applications~\cite{park97, ye98}. Neutron scattering has played a
particularly important role in the study of the unusual lattice
dynamics and diffuse scattering observed in single crystals of PMN
\cite{naberezhnov99,gehring01,wakimoto02-65,wakimoto02-66,vakhrushev02,
vakhrushev95,hirota02}.  Because of the $Q^{2}$-dependence of the
phonon and diffuse scattering cross sections, most studies to date
have made use of thermal neutrons ($\lambda\sim 2$~\AA) in order
to access Brillouin zones corresponding to large momentum transfer
$\mathbf{Q}$.   However, Xu \textit{et al.} have performed recent
experiments on PMN using cold neutron ($\lambda \geq 4$~\AA)
time-of-flight (TOF) techniques that have succeeded in observing
the diffuse scattering in the low-$\mathbf{Q}$ $(1,0,0)$ Brillouin
zone~\cite{xu03}. Other cold neutron studies have also reported
diffuse and low-energy TA phonon measurements made at
low-$\mathbf{Q}$~\cite{hlinka03-jpcm,gvasaliya03,gvasaliya04}.

In this paper we report cold neutron scattering experiments on PMN
using triple-axis spectroscopic techniques.  Our goal was to
explore the static and dynamic relaxor properties of PMN through
measurements of the diffuse scattering and the transverse-acoustic
(TA) phonons at small reduced wave-vector $\mathbf{q}$
in the $(1,0,0)$ and
$(1,1,0)$ zones.  The diffuse scattering in PMN is believed to be
directly associated with the formation of polar nanoregions (PNR)
at the Burns temperature ($T_d \sim 600$~K~\cite{burns83}), 
and it increases
rapidly with cooling~\cite{naberezhnov99,hirota02}.
This diffuse scattering keeps increasing through
the ferroelectric transition temperature 
($T_{C} \simeq 210$~K~\cite{Schmidt80,ye93,wakimoto02-65}).
Careful examination confirms the presence of a
very weak, but definitive, diffuse scattering cross section above
$T_d$.  The high-temperature contours of this scattering are quite
different in shape from those observed at low
temperatures~\cite{vakhrushev95,hirota02,xu03,you97,takesue01,chaabane03}.
These contours persist to lower temperatures, remaining
unchanged even after the much stronger low-temperature diffuse
scattering sets in.  The high-temperature contours are therefore
very likely related to the atomic shifts produced by the
underlying chemical short-range order intrinsic to PMN.

The use of cold-neutron wavelengths limits our measurements to the
$(1,0,0)$ and $(1,1,0)$ zones.  Moreover, the 
TA phonon measurements are only possible in the $(1,1,0)$ zone
because the dynamical structure factor of the TA phonon is nearly
zero at $(1,0,0)$ (see Table~\ref{Tab1}).  
A well-defined low-energy TA phonon mode is observed 
at the smallest $\mathbf{q}$, or $\zeta=0.035$, 
near $(1,1,0)$ over a wide range of temperatures, 
where $\mathbf{q}=(\zeta, -\zeta, 0)$.
During the course of our experiments,
Stock \textit{et al.} reported 
a dramatic line broadening of TA phonon 
for $T_{C} < T < T_d$ in the same $(1,1,0)$ zone at
$\zeta=0.1$ or higher~\cite{stock04}.
Therefore, we have extended our cold-neutron measurements up to 
$\zeta=0.1$ for temperatures around $T_d$ as well as 
below $T_{C}$.
Our results overlap and agree with their thermal neutron data.

\begin{table}[b]
\caption{Calculated structure factors for PMN. For
$|F_{soft}|^{2}$, the parameter $S_{2}/S_{1}$ is fixed to be 1.5
according to Ref.~\onlinecite{hirota02}. See the text for
$|F_{diff}|^{2}$. }
\begin{center}
\tabcolsep=1.5mm
\begin{tabular}{ccccc} \hline
\hspace{5mm} ---\hspace{5mm} & Bragg & TA phonon & Soft phonon & Diffuse\\
$(h k l)$ & $|F_{B}|^{2}$ & $Q^{2}|F_{B}|^{2}$ &
$Q^{2}|F_{soft}|^{2}$ & $Q^{2}|F_{diff}|^{2}$\\
\hline
$(1 0 0)$ & 1   & 0.2 & 11 & 11\\
$(1 1 0)$ & 9   & 5 & 9 & 32 \\
$(1 1 1)$ & 37  & 28 & 37 & 2 \\
$(2 0 0)$ & 100 & 100 & 91 & 0.1 \\
$(3 0 0)$ & 1 & 2 &100 & 100 \\
\hline
\end{tabular}
\end{center}
\label{Tab1}
\end{table}%

\section{EXPERIMENTAL DETAILS}

\begin{figure}[t]
\begin{center}
\includegraphics[scale=0.5,clip]{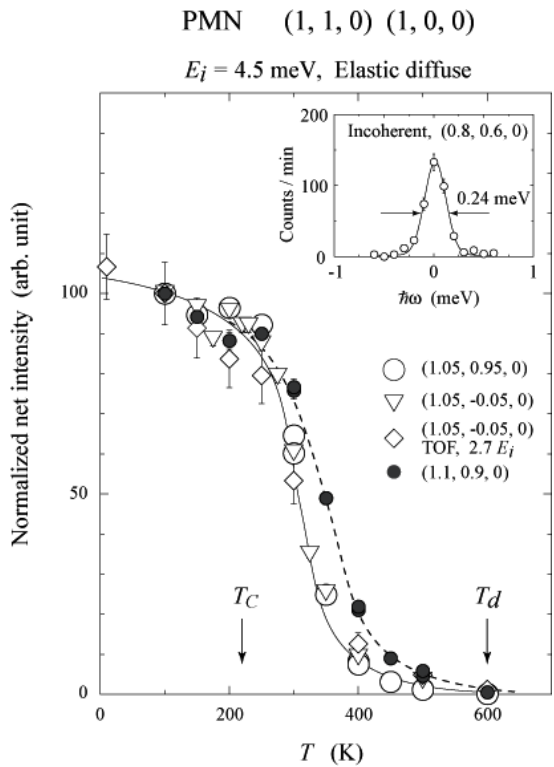} %\vskip 4pt
\caption{Temperature dependence of the elastic diffuse scattering
intensity measured 
at $\zeta=0.05$ and $0.1$. 
Data are normalized at 100~K.  The solid and broken
lines are guides to the eyes. 
Open diamonds show the results from TOF 
measurements~\cite{xu03}. 
}
\label{Fig3}
\end{center}
\end{figure}

Our experiments were carried out on the cold neutron triple-axis
spectrometer SPINS located at the NIST Center for Neutron
Research.  The PMN sample, grown at Simon Fraser University in
Canada, has a mass of 4.8~grams and is the same high-quality
single crystal that was studied in previous
reports~\cite{wakimoto02-65,wakimoto02-66}. The sample has a
room-temperature lattice parameter of $4.04$~\AA, so that 1~rlu
(reciprocal lattice unit) $=1.56$~\AA$^{-1}$.  The crystal [001]
axis was mounted vertically so that data were collected in the
$(h,k,0)$ scattering plane.  The initial (final) neutron energy
$E_{i}$ ($E_{f}$) was fixed at 4.5~meV ($\lambda=4.26$~\AA~and
$k=1.47$~\AA$^{-1}$) during elastic (inelastic) neutron
measurements, and a Be filter was placed before (after) the sample
to remove higher order neutrons.  The $(0,0,2)$ reflection of
highly-oriented pyrolytic graphite crystals was used to
monochromate and analyze the incident and scattered neutron
energies, respectively.  The beam collimations employed were
guide-80'-80'-open.  The measured instrumental energy resolution
for this configuration is $2\mathit{\Gamma} _{res} = 0.24$~meV as
shown in the inset of Fig.~\ref{Fig3}, and the background
level is about $4$~counts per minute.

\section{ELASTIC DIFFUSE SCATTERING}

\begin{figure}[t]
\begin{center}
\includegraphics[scale=0.65,clip]{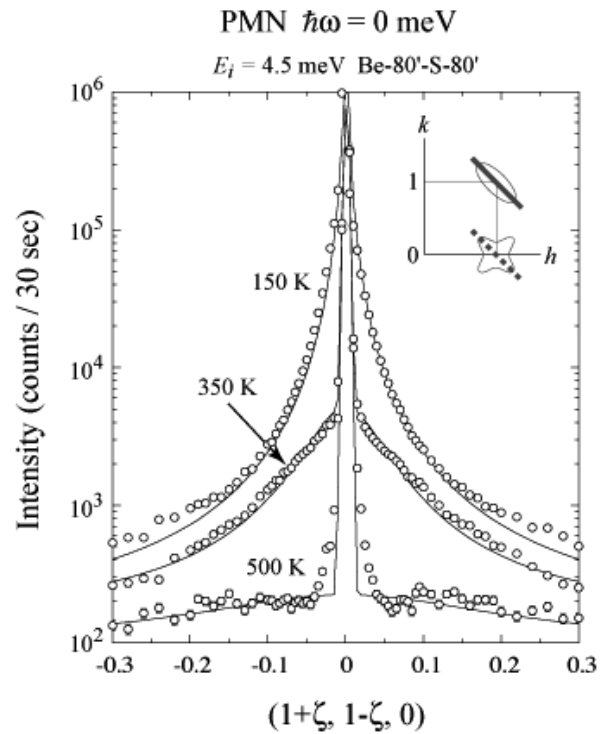} %\vskip 4pt
\caption{ Thermal evolution of the elastic diffuse scattering
around $(1,1,0)$ along the intensity ``ridge'' direction
(thick line, inset). 
}
\label{Fig2}
\end{center}
\end{figure}

Motivated by recent TOF measurements on PMN~\cite{xu03}, we have
performed a systematic investigation of the diffuse scattering
around both of the $(1,0,0)$ and $(1,1,0)$ Bragg peaks. 
The temperature dependence of the diffuse scattering intensity is
plotted in Fig.~\ref{Fig3} at several different
$\mathbf{Q}=\mathbf{G}+\mathbf{q}$ positions, where $\mathbf{G}$
represents either the $(1,1,0)$ or $(1,0,0)$ reciprocal lattice vector 
and $\mathbf{q}=(\zeta, -\zeta, 0)$.
All intensities have been normalized to 100 at 100~K. The
open-symbol data were measured at the same $\zeta=0.05$.
We first note that the
thermal evolution of the diffuse scattering intensity is very
similar around different Bragg peaks. Another interesting point is
that while the diffuse scattering intensity is visible just below
$T_{d}$, it grows rapidly below $\sim 400$~K.  This result is
in good agreement with previous high-resolution TOF measurements,
which are shown by the open diamonds in
Fig.~\ref{Fig3}~\cite{xu03}.  In the TOF study the instrumental
energy resolution ($E_{i} = 2.7$~meV and 
$2\mathit{\Gamma} _{res} =0.085$~meV)
was about three times better than that of the
SPINS experiment.  
The consistency between these independent
measurements is significant because it shows that the data do not
change with improving energy resolution.  Thus our data render an
accurate portrayal of the intrinsic temperature dependence of the
elastic diffuse scattering.  At larger $\mathbf{q}$, such as
$\zeta=0.1$, which is shown by the solid circles in
Fig.~\ref{Fig3}, the diffuse scattering intensity takes off at a
slightly higher temperature, so the whole curve (broken line) is
shifted toward higher temperatures. However the overall behavior
is still the same, i.e. the diffuse scattering intensity is always
first detected at or just below $T_d$, and increases monotonically
with cooling.

Typical $\mathbf{q}$-profiles of the elastic diffuse intensity
measured around the $(1,1,0)$ peak are shown in a semi-log plot in
Fig.~\ref{Fig2}.  The intensity profiles are well described by a
Gaussian function (Bragg peak) plus a Lorentzian function (diffuse
scattering) and a flat background.  The fits are shown as the
solid lines in Fig.~\ref{Fig2}.  With increasing temperature, the
intensity of the diffuse scattering decreases, and the width of
the Lorentzian increases, indicating a decrease in the correlation
length. This temperature variation agrees with that measured in
the $(1,0,0)$ zone
along the same $[1 \overline{1} 0]$ direction~\cite{xu03} 
(broken line, in the inset of Fig.~\ref{Fig2}).
Note that the diffuse intensity
becomes very weak even at 500~K, which is still 100~K below
$T_{d}$.

\begin{figure}[b]
\begin{center}
\includegraphics[scale=0.5,clip]{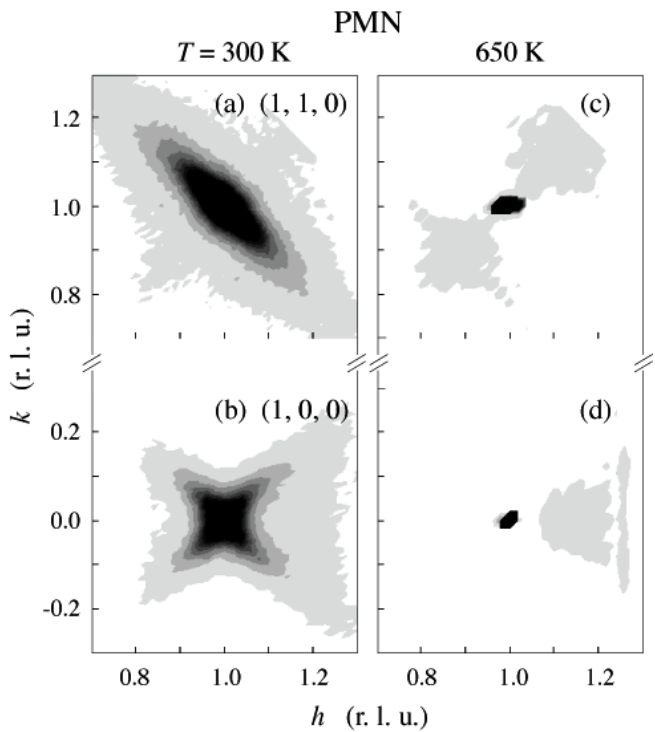} %\vskip 4pt
\caption{ Intensity contours of the elastic diffuse scattering
well below $T_{d}$ (left column) and just above $T_{d}$ (right
column), around the $(1,0,0)$ and $(1,1,0)$ Bragg peaks.}
\label{Fig1}
\end{center}
\end{figure}

Intensity contours of the diffuse scattering measured at 300~K and
650~K are shown in Fig.~\ref{Fig1}. At 300~K, which is above
$T_{C}$, but below $T_{d}$, the scattering is strong in both
zones, and in agreement with previous studies
~\cite{vakhrushev95,hirota02}.  The scattering intensity
forms a ``butterfly'' pattern around the (1,0,0) Bragg peak, with
ridges extending along the $[1~\pm 1~0]$ directions.  The 
scattering at $(1,1,0)$ is ellipsoidal in shape, and extends along
the $[1~\overline{1}~0]$ direction.  These results are consistent
with previous x-ray~\cite{you97,takesue01,chaabane03} and
neutron~\cite{vakhrushev95,hirota02,xu03} measurements.  On the
other hand, a weak pattern of diffuse scattering is still visible
around both of these Bragg peaks at 650~K, just above $T_{d}$
[Figs.~\ref{Fig1}~(c) and (d)].  This high-temperature scattering
exhibits a geometry that is quite different from that observed at
300~K. The intensity of the high-temperature diffuse scattering at
$(0.9, 0.9, 0)$ in Fig.~\ref{Fig1} (c), for example, is less than
$0.1$~\% of that of the low-temperature diffuse peak at $(1,1,0)$.

The presence of a diffuse cross section above $T_{d}$ is
very surprising. Therefore, we did an extra background 
check using a single crystal of SrTiO$_3$ ($\sim 2$ cc)
and the identical configuration as used with PMN.
No meaningful signal was observed near $(1,0,0)$ and $(1,1,0)$
within the experimental accuracy.

\begin{figure}[t]
\begin{center}
\includegraphics[scale=0.55,clip]{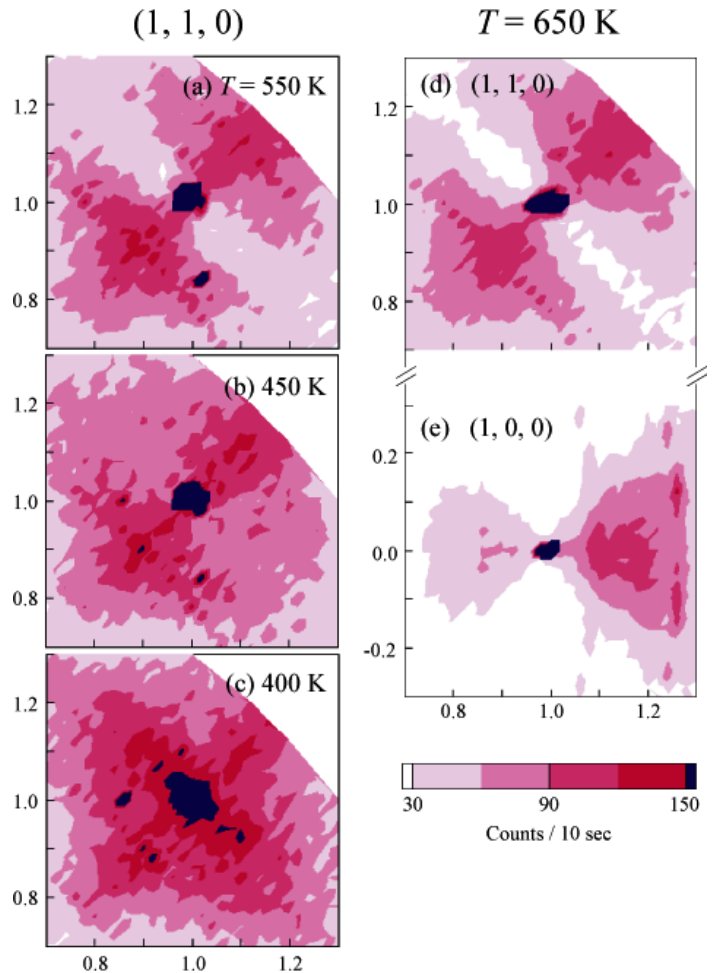} %\vskip 4pt
\caption{ Temperature variation of diffuse scattering contours
near $(1,1,0)$.
The background level is $\sim 20$~counts per 10~sec.
Tuned high-temperature contours measured at 650~K in (d) $(1,1,0)$
and (e) $(1,0,0)$ zones.}
\label{Fig7}
\end{center}
\end{figure}

In Fig.~\ref{Fig7}, the intensity scale was tuned to better illustrated  the 
contrast of the high-temperature diffuse contour (HTC).
The intensity level of the HTC is much weaker than that
of the low-temperature diffuse peak associated with the PNR
formation as listed in Table~\ref{Tab2}. However, because of
the clean window and the efficiency of the spectrometer, the signal
to noise ratio reaches about 4 even in the weak HTC. The HTC have
distinctly different shapes than the low temperature diffuse 
intensities in both $(1,1,0)$ and $(1,0,0)$ zones. They 
seem to peak at incommensurate positions around (1.1,0,0) and (0.9,0,0) in 
the $(1,0,0)$ zone, and (1.1,1.1,0) and (0.9,0.9,0) in the $(1,1,0)$ zone,
while the low temperature diffuse scattering intensities
always reach the maximum at the zone center $\mathbf{q}=0$. 
The intense spots at the center of the contour plots are scattering intensities
from the Bragg peak. On
cooling, around 450~K, the HTC can be clearly seen, superimposed
with the increasingly strong low temperature diffuse cross section [see
Figs.~\ref{Fig7} (a) - (c)]. The weak HTC are further seen down to
even lower temperatures in Figs.~\ref{Fig1}~(a) and (b), and
virtually unchanged below 650~K. 
Details on this high temperature diffuse 
cross section will be further discussed in the discussion section.

\begin{table}[t]
\caption{
Observed peak intensity (counts/min).
}
\begin{center}
\tabcolsep=5mm
\begin{tabular}{ccc} \hline
\hspace{5mm} ---\hspace{5mm} & $(1, 0, 0)$ & $(1,1,0)$\\
\hline
Bragg & $1.2\times 10^{6}$ & $2.0\times 10^{6}$ \\
100-K Diffuse &  $9\times 10^{5}$  & $6\times 10^{5}$ \\
650-K HTC & $500$   & $500$ \\
\hline
Incoherent & 130 & --- \\
Background & 4 & ---\\
\hline
\end{tabular}
\end{center}
\label{Tab2}
\end{table}%

\section{PHONON AND BROAD DIFFUSE SCATTERING}

\begin{figure}[b]
\begin{center}
\includegraphics[scale=0.6,clip]{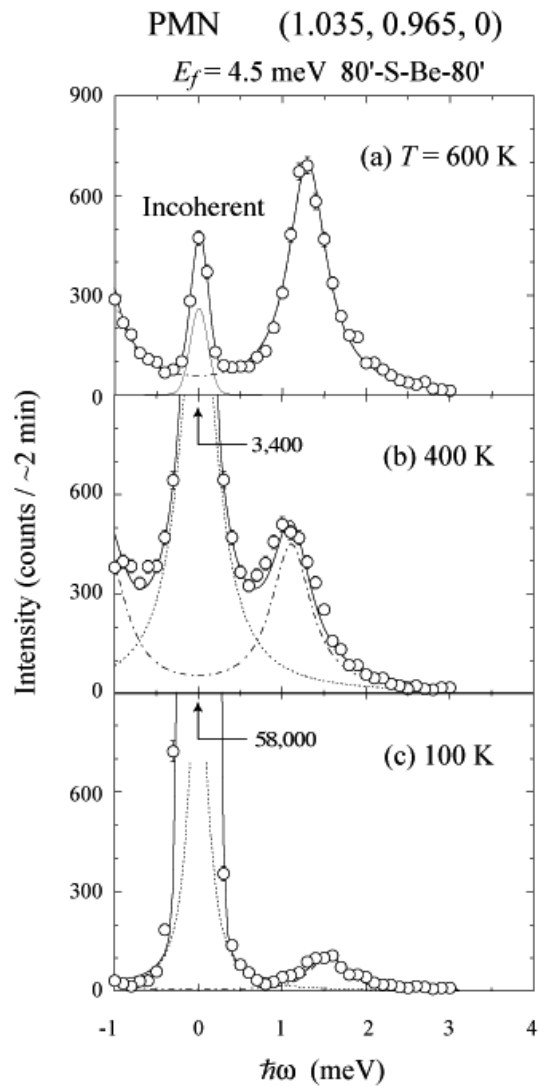} %\vskip 4pt
\caption{ Temperature variation of constant-$\mathbf{Q}$ scans
measured at small $\mathbf{q}$ near (1,1,0).  Fits are described
in the text. } \label{Fig4}
\end{center}
\end{figure}

\begin{figure}[t]
\begin{center}
\includegraphics[scale=0.6,clip]{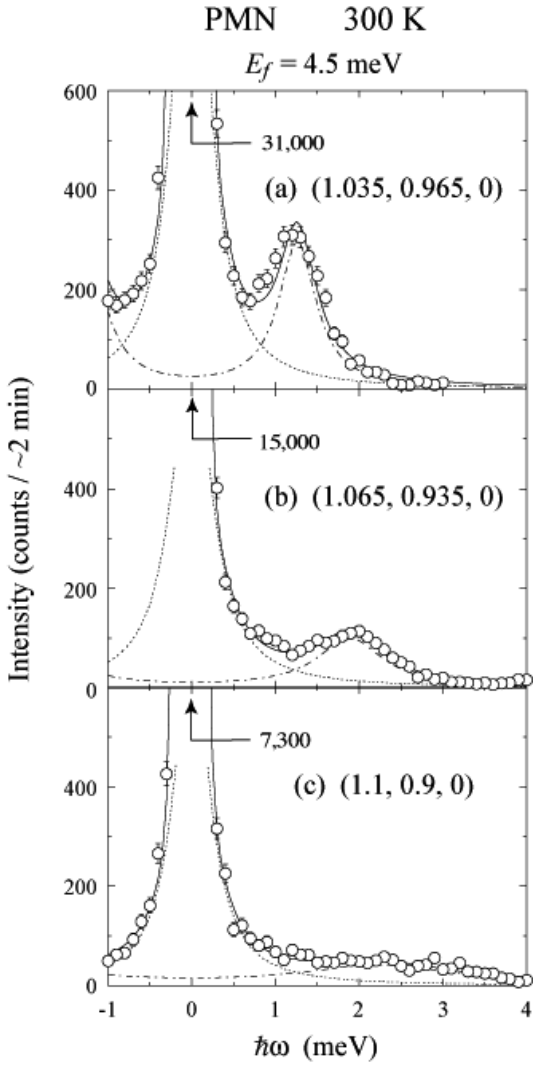} %\vskip 4pt
\caption{ $\mathbf{q}$ variation of constant-$\mathbf{Q}$ scans
measured at 300~K near (1,1,0). } \label{Fig5}
\end{center}
\end{figure}

\begin{figure}[t]
\begin{center}
\includegraphics[scale=0.6,clip]{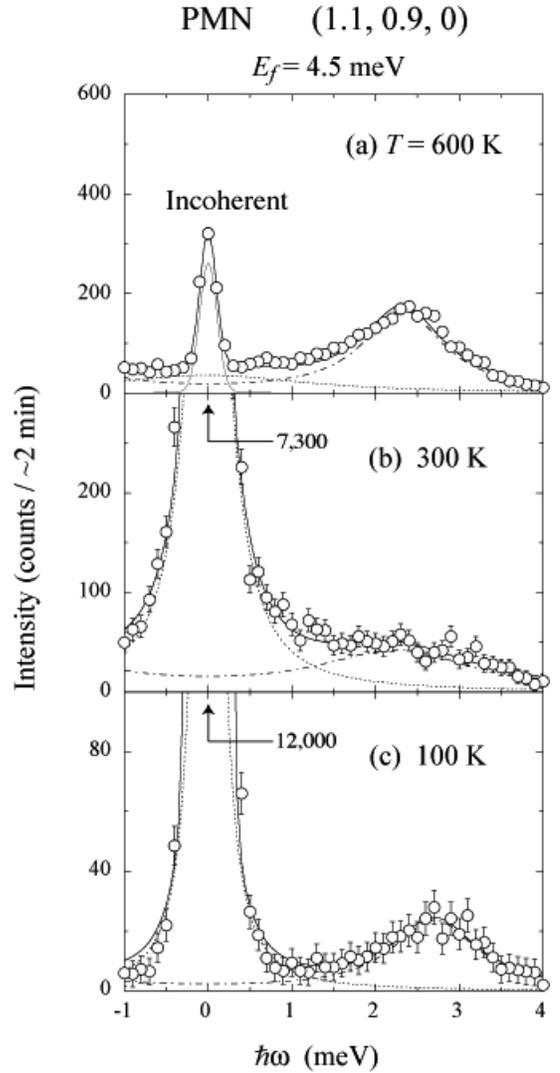} %\vskip 4pt
\caption{
Constant-$\mathbf{Q}$ scans at $\zeta=0.1$ for three temperatures.
The incoherent scattering is shown by a thin line only in (a).
}
\label{Fig8}
\end{center}
\end{figure}

To date the transverse acoustic (TA) phonons in PMN have mainly
been studied using thermal neutrons.  
All measurements are in good agreement in the $(2,0,0)$ 
and $(2,2,0)$ zones~\cite{naberezhnov99,gehring01,wakimoto02-65,
wakimoto02-66} where diffuse scattering is very weak
(see Table~\ref{Tab1}).
Naberezhnov \textit{et al}.~\cite{naberezhnov99} discovered 
that the TA-phonon starts to broaden at $T_{d}$.
Later on, Wakimoto {\it et al.}\cite{wakimoto02-65} found
this slight broadening disappears at $T_{C}$.
However, 
pronounced TA-phonon anomalies were observed in $(3,0,0)$
and $(1,2,2)$ zones~\cite{naberezhnov99,wakimoto02-66,
vakhrushev02},
where the diffuse cross section is large 
(see Table~\ref{Tab1}).
These anomalies were interpreted as the results of
different types of phonon interactions. 
Our current measurements are limited in small $\mathbf{Q}$ zones 
as a result of the scattering geometry for cold neutrons,
and are centered around $(1,1,0)$.
$(1,1,0)$ is a good zone to study both TA phonon and
diffuse scattering,
because interaction from the soft transverse-optic (TO) phonon 
must be very weak due its small cross section.

The fine energy resolution intrinsic to cold neutron experiments
($E_{f}\sim 4$~meV and $2\mathit{\Gamma} _{res} \sim 0.2$~meV for
typical triple-axis measurements) has the great advantage of
enabling the measurement of low-energy excitations at small
$\mathbf{q}$. To clarify the relationship between the 
diffuse scattering and the TA phonon in PMN, the
dynamic scattering function $S(\mathbf{q}, \omega)$ was studied
along the strong diffuse intensity ridge 
$\mathbf{q }= (\zeta, -\zeta, 0)$ 
for $0.035 \leq \zeta \leq 0.1$. The TA-phonon structure factor at
$(1,1,0)$ is much larger than that at $(1,0,0)$, as demonstrated
in Table~\ref{Tab1}.  The diffuse scattering structure factor
$Q^{2}|F_{diff}|^{2}$ listed there is calculated from the atomic
shifts derived from neutron diffuse scattering measurements by
Vakhrushev \textit{et al.}~\cite{vakhrushev95}

Energy spectra measured at $\zeta =0.035$
for several different temperatures are shown in
Fig.~\ref{Fig4}.  A well-defined TA phonon mode is observed at the
smallest $\mathbf{q}$ over a wide range of temperatures in spite of
the relatively small value of $\mathbf{Q}$ and the small
resolution volume.  At 400~K the energy spectrum shows an
interesting and definite overlap of the TA phonon with the central diffuse 
scattering intensity. This overlap could be the result of a softening of the
TA phonon energy and the prominent tail of the elastic central peak.
This suggests that extra spectral weight exists in the central 
peak in addition to the resolution-limited elastic diffuse intensity.
A small portion of the
diffuse tail still remains at 100~K, but completely disappears at
600~K. From these data we conclude that no elastic diffuse
scattering is present at 600~K at this $\mathbf{q}$.  

Energy spectra measured at 300~K and for $\mathbf{q}$ up to $\zeta =
0.1$ are shown in Fig.~\ref{Fig5}.  The overlap of the TA
phonon with the broad central peak is still
evident in this region.  It is interesting to note that the TA
phonon broadens markedly with increasing $\mathbf{q}$ at 300~K, as
is quite apparent at $\zeta=0.1$.
In order to overlap with the thermal neutron measurements
of Stock \textit{et al.}~\cite{stock04},
we have carried out further measurements at $\zeta=0.1$
for 600~K ($\sim T_{d}$) as well as 100~K ($\ll T_{C}$).
Figure~\ref{Fig8} traces the evolution of the
TA phonon at $(1.1, 0.9, 0)$.
The broadening reaches a climax at the
intermediate temperature 300~K. At $T=100$~K, which is well below $T_{C}$,
the TA phonon at $\zeta=0.1$ almost recovers its original line shape.

In the course of the current data analyses for energy spectra typically shown
in Figs.~\ref{Fig4} - \ref{Fig8},
we used the following scattering function to fit :
\begin{equation}
S(\mathbf{q}, \omega) = S_{rlD} + S_{brD} + S_{TA}.
\label{eq1}
\end{equation}
Here, $S_{rlD}$ represents the resolution-limited elastic diffuse peak
and it is expressed as a Gaussian function with the full width of 
$2\mathit{\Gamma} _{res}$
as previously done~\cite{naberezhnov99,vakhrushev02,gvasaliya03,gvasaliya04}.
$S_{brD}$ is the broad diffuse scattering prominent for intermediate
temperatures $T_{C}<T<T_{d}$.
$S_{TA}$ describes the TA phonon scattering.
We model $S_{brD}$ and $S_{TA}$ in the following way:
\begin{eqnarray}
S_{brD} \sim \frac{\mathit{\Gamma} _{diff}}
{\omega^2+\mathit{\Gamma} _{diff}^2},
\label{eq2}\\
S_{TA} \sim \frac{\mathit{\Gamma} _{TA}}
{(\omega-\omega _{TA})^2+\mathit{\Gamma} _{TA}^2}.
\label{eq3}
\end{eqnarray}
In this model
we regard the broad diffuse scattering as a part of 
intrinsic cross section for PNR.

These functional forms in Eqs.~(\ref{eq1}) - (\ref{eq3})
fit the data quite well as shown by the solid lines in
Figs.~\ref{Fig4} - \ref{Fig8}, where the phonon and the
broad central diffuse component are depicted by the chained and
dotted lines, respectively.  
Fitting parameters at $\zeta=0.035$ and $0.1$ for three temperatures
are listed in Table~\ref{Tab3}.
$|F_{TA}|$ is derived from the scale factor for Eq.~\ref{eq2}
after removing the prefactor of phonon intensity $(T/\omega _{TA}^2)$.
The peculiar broadening of the TA phonon is supported numerically 
in $\mathit{\Gamma} _{TA}$ at $\zeta=0.1$ for $T_{C}<T<T_{d}$.
In contrast, the TA-phonon mode at $\zeta= 0.035$ is well behaved
for a wide range of temperatures
with the nearly temperature-independent linewidth
and the dynamical structure factor.

\begin{figure}[t]
\begin{center}
\includegraphics[scale=0.55,clip]{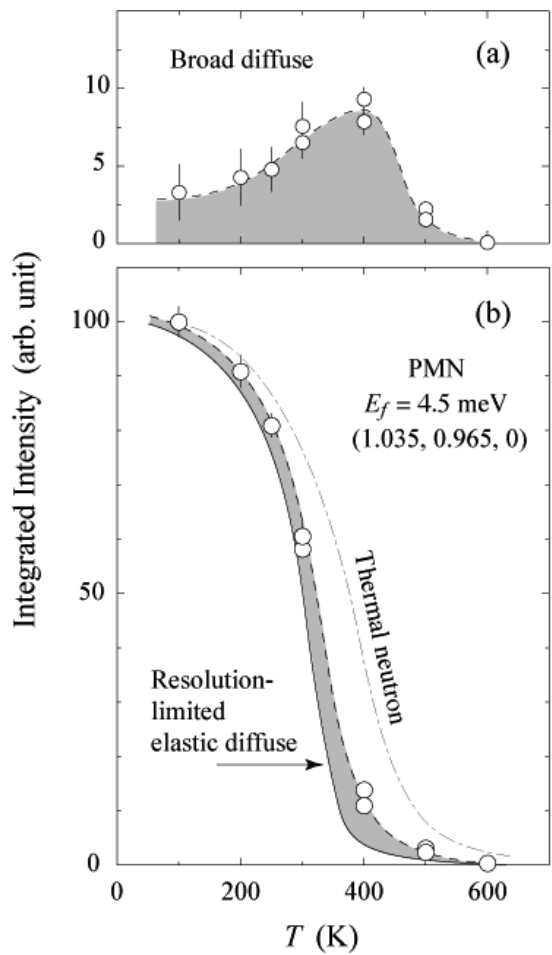} %\vskip 4pt
\caption{ The energy-integrated intensity of the diffuse
scattering as a function of temperature. The shaded region
represents the contribution from the broad central component.
The resolution-limited elastic component is shown by the solid line. 
All curves are guides to the eye. } 
\label{Fig6}
\end{center}
\end{figure}

\begin{table}[t]
\caption{
Representative fitting parameters
without resolution corrections.
$I_{diff}$ and $|F_{TA}|$ are related to the scale factors
in Eqs.~(\ref{eq2}) and (\ref{eq3}), respectively.
}
\begin{center}
\tabcolsep=2mm
\begin{tabular}{ccccccc} \hline
$\zeta$ & $T$ & $\omega _{TA}$ & $|F_{TA}|$ &  
$2\mathit{\Gamma} _{TA}$ & $2\mathit{\Gamma }_{diff}$ & $I_{diff}$ \\
(rlu) & (K) & (meV) & --- & (meV) & (meV) & --- \\
\hline
0.035 & 100 & 1.6 & 100 & 0.6 & 0.3 & 3 \\
0.035 & 300 & 1.2 & 80 & 0.6 & 0.4 & 8 \\
0.035 & 600 & 1.3 & 90 & 0.5 & --- & 0 \\
%0.035 & 100 & 1.6 & 100 & 0.3 & 0.1 & 3 \\
%0.035 & 300 & 1.2 & 61 & 0.3 & 0.2 & 8 \\
%0.035 & 600 & 1.3 & 77 & 0.3 & --- & 0 \\
\hline
0.1 & 100 & 2.7 & 150 & 1.6 & 0.3 & 7 \\
0.1 & 300 & 2.3 & 110 & 2.4 & 0.5 & 19 \\
0.1& 600 & 2.3 & 110 & 1.2 & 2.8 & 6 \\
%0.1 & 100 & 2.7 & 220 & 0.8 & 0.1 & 7 \\
%0.1 & 300 & 2.3 & 130 & 1.2 & 0.2 & 19 \\
%0.1& 600 & 2.3 & 120 & 0.6 & 1.4 & 6 \\
\hline
\end{tabular}
\end{center}
\label{Tab3}
\end{table}%

The temperature dependence of the energy-integrated intensity
of the diffuse scattering at $\zeta=0.035$ ($I_{diff}$ in Table~\ref{Tab3}) 
is shown in Fig.~\ref{Fig6}. 
The broad central component measured at the
smallest $\mathbf{q}$ peaks around 400~K, reaching about 10~\% of
the low-temperature resolution-limited elastic value
(shaded region in Fig.~\ref{Fig6}). 
Below 400~K, which is far below $T_{d}$, 
the sharp elastic component
grows quickly and becomes dominant
[see Fig.~\ref{Fig6}~(b)]. As reported in previous
papers~\cite{naberezhnov99,hirota02}, the diffuse scattering
intensity, corresponding to the formation of PNR, begins to
develop around $T_{d}$ when observed with thermal neutrons as
schematically shown in Fig.~\ref{Fig6}~(b). The difference between the
thermal and cold neutron results can then be naturally attributed
to differences in the experimental resolution functions.

\section{DISCUSSION}

The most important new result of the current experiment is
the observation of the HTC in Figs.~\ref{Fig7} (d) and (e).
The scattering pattern is longitudinal to $\mathbf{Q}$.
Also the intensity at larger $\mathbf{Q}$ is
stronger than that at smaller $\mathbf{Q}$ [Fig.~\ref{Fig7} (e)],
consistent with diffuse scattering resulting from atomic shifts,
where $I\propto |{\bf Q}\cdot\epsilon|^2$. Theoretical
studies~\cite{Burton01,Burton02} have suggested that short-range
chemical order between Mg$^{2+}$ and Nb$^{5+}$ ions at the
perovskite $B$-sites should generate local electric field that
leads to shifts of the $A$-site Pb$^{2+}$ ions. This short-range
chemical order is believed to be temperature insensitive up to $T
\sim 1000$~K. We speculate that this broad incommensurate cross
section in the HTC reflects atomic shifts that modulate in the
longitudinal direction, caused by the short-range chemical order.
These atomic shifts are also short-range correlated, resulting in
the diffuse type scattering pattern around the incommensurate
positions. Finally, the shape of the HTC looks somewhat conjugate
with that of the low-temperature strong diffuse cross section.
This may be more than merely a coincidence. It is likely that
the diffuse contours from the PNR grow out of the HTC with
cooling, but this is just a conjecture at present.

In the diffuse-phonon-profile analysis of 
the $(1,1,0)$-zone data,
the additional broad diffuse cross section of $S_{brD}$ was interpreted
as the part of intrinsic dynamics of PNR.
It is the same approach employed by 
Gvasaliya \textit{et al.}~\cite{gvasaliya03,gvasaliya04}
This broad Lorentzian cross section may contain
intensity transferred from the TA phonon by coupling to the 
resolution-limited elastic peak~\cite{Shapiro72}.
So far we could not obtain a unique set of parameters by
such the mode coupling.
The cold neutron results  reported here and by others
~\cite{gvasaliya03,gvasaliya04}
appear to provide a complete picture of the low-$\mathbf{q}$ scattering 
near $(1,1,0)$ and $(1,0,0)$.
However, previous thermal-neutron measurements around $(3,0,0)$ and 
$(1,2,2)$ reported complex cross sections with conflicting interpretations
~\cite{naberezhnov99,wakimoto02-66,vakhrushev02,hlinka03-prl}.
Thus, further neutron measurements, with high resolution in both
$(1,0,0)$ and $(3,0,0)$ zones, are vital for the 
understanding of the PNR dynamics in PMN.

\vspace{5mm}

%\newpage
\begin{acknowledgments}


We are very grateful to C.~Stock for sharing his PMN data with us
before publication.  We also thank B.~P.~Burton, E.~J.~Cockayne,
S. Prosandeev, S.~M.~Shapiro, S.~B.~Vakhrushev, and D.~Viehland
for stimulating discussions.  This study was supported by the
U.S.-Japan Cooperative Neutron-Scattering Program.  Financial
support from the U.S. Department of Energy under Contract
DE-AC02-98CH10886 is also gratefully acknowledged.  Work at SPINS
is based upon activities supported by the NSF under DMR-9986442.
We also acknowledge the U.S. Dept. of Commerce, NIST Center for
Neutron Research, for providing the neutron scattering facilities
used in this study.

\end{acknowledgments}

\bibliography{basename of .bib file}% // USE this command later.

\end{document}